\RequirePackage{fix-cm}
\documentclass[twocolumn,final]{svjour3}          
\smartqed  
\usepackage{graphicx}
\journalname{Brazilian Journal of Physics}
\begin{document}

\title{Coupling modifies the quantum fluctuations of entangled oscillators
}


\author{Roberto Baginski B. Santos         \and
        Vinicius S. F. Lisboa 
}


\institute{R. B. B. Santos \at
              Centro Universit\'{a}rio FEI \\
              \email{rsantos@fei.edu.br}           
           \and
           V. S. F. Lisboa \at
              Universidade Federal do ABC \\
              \email{vinicius.lisboa@ufabc.edu.br}
}

\date{Received: date / Accepted: date}

\maketitle

\begin{abstract}
Coupled oscillators are among the simplest composite quantum systems in which the interplay of entanglement and interaction may be explored. We examine the effects of coupling on fluctuations of the coordinates and momenta of the oscillators in a single-excitation entangled state. We discover that coupling acts as a mechanism for noise transfer between one pair of coordinate and momentum and another. Through this noise transfer mechanism, the uncertainty product is lowered, on average, relatively to its non-coupled level for one pair of coordinate and momentum and it is enhanced for the other pair. This novel mechanism may be explored in precision measurements in entanglement-assisted sensing and metrology.
\end{abstract}

\section{Introduction}
\label{intro}

Entanglement and interaction are defining features of composite quantum systems. Entanglement is at the heart of quantum mechanics \cite{Zeilinger:Phys_Scr_1998,Horodecki:RMP_2009,Peschel:BJP_2012,Aolita+de_Melo+Davidovich:Rep_Prog_Phys_2015}, being recognized as an important resource for the technologies of the second quantum revolution \cite{Dowling+Milburn:PTRSLA_2003} such as computation \cite{Jozsa:PRSLA_1998,Ekert+Jozsa:PTRSLA_1998,Steane:Rep_Prog_Phys_1998,Bennett+DiVincenzo:Nature_2000,Bouwmeester+Ekert+Zeilinger:PQI_2000,Nielsen+Chuang:QCQI_2000,Jozsa+Linden:PRSLA_2003,Mermin:QCS_2007,Preskill:Quantum_2018}, quantum communication and cryptography \cite{Ekert:PRL_1991,Bennett+Wiesner:PRL_1992,Bennett+etal:PRL_1993,Zukowski+etal:PRL_1993,Gisin+etal:RMP_2002,Gisin+Thew:Nature_Photonics_2007,Broadbent+Schaffner:Designs_Codes_Crypto_2016,Diamanti+etal:npjQI_2016,Yin+etal:Nature_2020,Pirandola+etal:Adv_Opt_Photon_2020}, and sensing and metrology \cite{Bollinger+etal:PRA_1996,Giovannetti+Lloyd+Maccone:Science_2004,Giovannetti+Lloyd+Maccone:PRL_2006,Fernholz+etal:PRL_2008,Wasilewski+etal:PRL_2010,Giovanetti+Lloyd+Maccone:Nature_Photonics_2011,Degen+Reinhard+Cappellaro:RMP_2017}.

In many cases, the parts of a composite system interact. Sometimes, a strong coupling between the parts of a system is desired in order to enable the observation of certain phenomena, especially those involving just a few excitations in cavity QED \cite{Rempe+Walther+Klein:PRL_1987,Thompson+Rempe+Kimble:PRL_1992,Brune+etal:PRL_1996a,Kimble:Phys_Scr_1998,Raimond+Brune+Haroche:RMP_2001,Walther+etal:Rep_Prog_Phys_2006} or in other systems \cite{Bouchiat+etal:Phys_Scr_1998,Nakamura+Pashkin+Tsai:Nature_1999,Leibfried+etal:RMP_2003,Chiorescu+etal:Nature_2004,Wallraff+etal:Nature_2004,Blais+etal:PRA_2004,Reithmaier+etal:Nature_2004,Clarke+Wilhelm:Nature_2008,Groblacher+etal:Nature_2009,Niemczyk+etal:Nature_Physics_2010,Teufel+etal:Nature_2011} that are being explored more recently such as circuit QED, quantum dots, optomechanical resonators and semiconductor microcavities, for instance. In the strong coupling regime, it is possible to observe reversibility of the coherent exchanges between the parts of the system.

On the other hand, the interaction of a system and its environment is responsible for dissipation and decoherence effects that wipe away energy and quantum correlations and lead to irreversibility and the emergence of the classical behavior \cite{Zeh:FP_1970,Zurek:PRD_1981,Zurek:PRD_1982,Caldeira+Leggett:Physica_1983,Caldeira+Leggett:Ann_Phys_1983,Joos+Zeh:ZPB_1985,Caldeira+Leggett:PRA_1985,Unruh+Zurek:PRD_1989,Brune+etal:PRL_1996b,Zurek:RMP_2003,Joos+etal:DACWQT_2003,Omnes:BJP_2005,Jacquod+Petitjean:Adv_in_Phys_2009}.

A few recent papers analyzed the dynamics of entanglement in coupled oscillators \cite{Paz+Roncaglia:PRL_2008,Kao+Chou:NJP_2016,Makarov:PRE_2018}. It was shown that a separable state evolves into an entangled state under the influence of the interaction between the oscillators, revealing features as death and revival of entanglement.

In this paper, we investigate the effect of coupling on coordinate and momentum fluctuations of two entangled coupled oscillators.

We present closed form expressions for the time evolution of the fluctuations, discuss the strong and the very strong coupling regimes and show that coupling leads to a transfer of noise among one pair of coordinate and momentum and another. Understanding how coupling affects the fluctuations of entangled oscillators may be useful in areas such as quantum sensing and metrology.

\section{The model}
\label{sec:model}

We consider a system composed of two identical oscillators with natural frequency $\omega$ and position-position coupling with strength $\Omega$. The Hamiltonian describing this system is \cite{Ester+Keil+Narducci:PR_1968,Han+Kim+Noz:AmJPhys_1999}
\begin{equation}\label{eq:H_coupled_1}
 H = \frac{1}{2}\left(\sum_i \left( p_i^2 + \omega^2 x_i^2\right) + \Omega^2\left(x_1 - x_2 \right)^2 \right),
\end{equation}
where $x_i$, $p_i$ are coordinate and momentum operators associated with oscillator $i$. As usual, these operators satisfy commutation relations $[x_i,p_j]=i\hbar\delta_{ij}$. Very recently, coupled quantum oscillators were used to model light propagation in an inhomogeneous medium \cite{Urzua+etal:SciRep_2019}. Parametric coupled oscillators were also analyzed \cite{Macedo+Guedes:J_Math_Phys_2012,Urzua-Pineda+etal:Q_Rep_2019}.

This Hamiltonian may be easily cast in the alternative form
\begin{equation}\label{eq:H_coupled_2}
 H = \frac{1}{2}\left(\sum_i \left( p_i^2 + \omega'^2 x_i^2\right) - 2\Omega^2 x_1x_2 \right),
\end{equation}
with $\omega'^2 = \omega^2 + \Omega^2$, which highlights the fact that the coupling is of the position-position type. We will not resort to the resonant or rotating-wave approximation, since results obtained with the resonant approximation are accurate only in the weak coupling case \cite{Ester+Keil+Narducci:PR_1968}, in which $\Omega/\omega\ll 1$.

As it is well known, incompatibility between coordinate and momentum operators leads to fluctuations. For a single oscillator in a number state with energy $E_n=(2n+1)\hbar\omega/2$, coordinate fluctuation is given by
\begin{equation}
 \Delta_n x = \sqrt{(2n+1)}\sqrt{\frac{\hbar}{2\omega}} = \sqrt{2n+1}\Delta_0 x
\end{equation}
while momentum fluctuation is
\begin{equation}
 \Delta_n p = \sqrt{(2n+1)}\sqrt{\frac{\hbar\omega}{2}} = \sqrt{2n+1}\Delta_0 p
\end{equation}
where the fluctuation amplitude $\Delta_\Gamma Y (t)$ of the operator $Y(t)$ in a state $|\Gamma\rangle$ is measured by the standard deviation
\begin{equation}
 \Delta_\Gamma Y (t) = \sqrt{\langle\Gamma | Y^2 (t) | \Gamma\rangle - \langle\Gamma | Y(t) | \Gamma\rangle^2}.
\end{equation}
It is worth remembering that the product of coordinate and momentum fluctuations in a number state is $\Delta_n x \Delta_n p = (2n+1)\hbar/2$. Squeezing may diminish coordinate fluctuations while increasing momentum fluctuations or vice-versa \cite{Caves:PRD_1981,Caves+Schumaker:PRA_1985,Schumaker+Caves:PRA_1985}, but it may not reduce the product of fluctuations below the Heisenberg limit \cite{Heisenberg:ZP_1927,Kennard:ZP_1927,Condon:Science_1929,Robertson:PR_1929,Heisenberg:PPQT_1930}.

When the oscillators are coupled, we may introduce normal mode operators for coordinates
\begin{equation}\label{eq:X_def}
 X_{\pm} = \frac{1}{\sqrt{2}}(x_1 \pm x_2),
\end{equation}
and for momenta
\begin{equation}\label{eq:P_def}
 P_{\pm} = \frac{1}{\sqrt{2}}(p_1 \pm p_2).
\end{equation}
These operators form a canonically conjugate pair satisfying $[X_\alpha, P_\beta] = i\hbar\delta_{\alpha\beta}$. With these normal mode operators, the Hamiltonian decouples, becoming
\begin{equation}\label{eq:H_decoupled}
 H = \frac{1}{2}\sum_\alpha \left( P_\alpha^2 + \omega_\alpha^2 X_\alpha^2\right)
\end{equation}
where the slow mode ($\alpha=+$) pulses with angular frequency $\omega_+=\omega$, and the fast mode ($\alpha=-$) pulses with angular frequency
\begin{equation}
 \omega_- = \omega\sqrt{1+2\frac{\Omega^2}{\omega^2}} = \eta\omega \geq \omega_+,
\end{equation}
and the factor $\eta$ is defined by
\begin{equation}
 \eta = \sqrt{1+2\frac{\Omega^2}{\omega^2}} \geq 1.
\end{equation}
For classical coupled oscillators, the slow mode corresponds to the symmetrical, in-phase normal mode, and the fast mode is the antisymmetrical, out-of-phase normal mode.

Using Heinsenberg equation of motion
\begin{equation}
 \dot{Y} = \frac{1}{i\hbar}[Y,H],
\end{equation}
it is simple to obtain closed-form expressions for the normal mode coordinate operators
\begin{equation}
 X_\pm (t) = X_\pm (0)\cos(\omega_\pm t) + \frac{P_\pm (0)}{\omega_\pm}\sin(\omega_\pm t)
\end{equation}
and for the momentum operators
\begin{equation}
 P_\pm (t) = P_\pm (0)\cos(\omega_\pm t) - \omega_\pm P_\pm (0)\sin(\omega_\pm t).
\end{equation}

Introduction of the ladder operators
\begin{eqnarray}
 A_\pm &=& \sqrt{\frac{\omega_\pm}{2\hbar}}\left(X_\pm + \frac{i}{\omega_\pm}P_\pm \right) \\
 A_\pm^\dagger &=& \sqrt{\frac{\omega_\pm}{2\hbar}}\left(X_\pm - \frac{i}{\omega_\pm}P_\pm \right)
\end{eqnarray}
satisfying $[A_\pm,A\pm^\dagger]=1$ eases the determination of matrix elements involving the operators $X_\pm$, $P_\pm$, their products and powers, since
\begin{eqnarray}
 X_\pm &=& \sqrt{\frac{\hbar}{2\omega_\pm}}\left(A_\pm^\dagger + A_\pm\right) \\
 P_\pm &=& i\sqrt{\frac{\hbar\omega_\pm}{2}}\left(A_\pm^\dagger - A_\pm\right)
\end{eqnarray}
and
\begin{eqnarray}
 A_\pm |N_\pm \rangle &=& \sqrt{N} |N_\pm - 1\rangle \\
 A_\pm^\dagger |N_\pm \rangle &=& \sqrt{N+1} |N_\pm + 1\rangle.
\end{eqnarray}

Inverting equations \ref{eq:X_def} and \ref{eq:P_def}, we obtain
\begin{eqnarray}
\label{eq:x_1}\nonumber
 x_1(t) &=& \sum_{\alpha}\frac{1}{\sqrt{2}}\Big( X_\alpha(0)\cos(\omega_\alpha t) \\ && + \frac{P_\alpha(0)}{\omega_\alpha}\sin(\omega_\alpha t) \Big) \\
 \label{eq:x_2}\nonumber
 x_2(t) &=& \sum_{\alpha}\frac{(-1)^\alpha}{\sqrt{2}}\Big( X_\alpha(0)\cos(\omega_\alpha t) \\  && +  \frac{P_\alpha(0)}{\omega_\alpha}\sin(\omega_\alpha t) \Big) \\
 \label{eq:p_1}\nonumber
 p_1(t) &=& \sum_{\alpha}\frac{1}{\sqrt{2}}\Big( P_\alpha(0)\cos(\omega_\alpha t) \\ && - \omega_\alpha X_\alpha(0)\sin(\omega_\alpha t) \Big) \\
 \label{eq:p_2}\nonumber
 p_2(t) &=& \sum_{\alpha}\frac{(-1)^\alpha}{\sqrt{2}}\Big( P_\alpha(0)\cos(\omega_\alpha t) \\ && - \omega_\alpha X_\alpha(0)\sin(\omega_\alpha t) \Big)
\end{eqnarray}

The entangled states considered in this paper are Bell-like states in the single-excitation regime:
\begin{equation}
 |\Psi^\pm\rangle = \frac{1}{\sqrt{2}}(|01\rangle \pm |10\rangle).
\end{equation}
With photons, for instance, such entangled two-mode Fock states may be generated using spontaneous parametric down-conversion processes in a $\chi^{(2)}$ nonlinear crystal with postselection \cite{Resch+Lundeen+Steinberg:PRL_2002,Banaszek+etal:PRA_2002} or with a heralded scheme \cite{Gubarev+etal:PRA_2020}. Entanglement with mechanical oscillators has also been demonstrated with trapped ions \cite{Jost+etal:Nature_2009}, with electromechanical circuits \cite{Palomaki+etal:Science_2013}, and, more recently, with micromechanical drum oscillators coupled via a superconducting circuit \cite{Ockeloen-Korppi+etal:Nature_2018}.

\section{Results}
\label{sec:results}

In order to determine coordinates and momenta fluctuation amplitudes, we need to evaluate a few matrix elements of the form $\langle \Psi^\pm | Y | \Psi^\pm \rangle$ between the entangled states. For easy of reference, these matrix elements are given in table \ref{tab:mat_elem}.

\begin{table}
\caption{Matrix elements for normal mode operators between Bell-like entangled states. In the entries, $\alpha\neq\beta$.}
\label{tab:mat_elem}
\begin{tabular}{ll}
\hline\noalign{\smallskip}
Operator $Y$ & $\langle \Psi^\pm | Y | \Psi^\pm \rangle$ \\
\noalign{\smallskip}\hline\noalign{\smallskip}
 $X_\alpha(0)$ & $0$ \\
 $P_\alpha(0)$ & $0$ \\
 $X_\alpha^2(0)$ & $\hbar/\omega_\alpha$ \\
 $P_\alpha^2(0)$ & $\hbar\omega_\alpha$ \\
 $X_\alpha(0) X_\beta(0)$ & $\pm\hbar/2\sqrt{\eta}\omega$ \\
 $X_\alpha(0) P_\alpha(0)$ & $i\hbar/2$ \\
 $P_\alpha(0) X_\alpha(0)$ & $-i\hbar/2$ \\
 $P_\alpha(0) P_\beta(0)$ & $\displaystyle \pm\sqrt{\eta}\hbar/2\omega$ \\
\noalign{\smallskip}\hline
\end{tabular}
\end{table}

\begin{figure}
 \includegraphics[width=\columnwidth,bb = 134 414 478 657,clip]{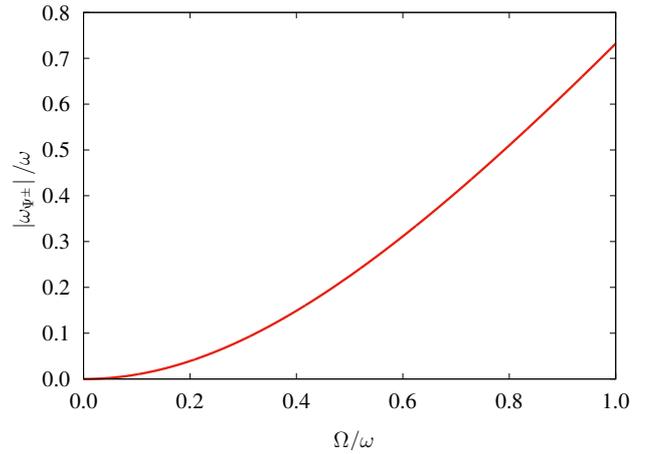}
 \caption{Relative frequency $|\omega_{\Psi^\pm}|/\omega$ as a function of the coupling strength $\Omega/\omega$.}
\label{fig:frequency_shift}
\end{figure}

With the matrix elements displayed in table \ref{tab:mat_elem} and equations \ref{eq:x_1}-\ref{eq:p_2} for the coordinates $x_1(t), x_2(t)$ and for the momenta $p_1(t), p_2(t)$, it is possible to determine normalized coordinates and momenta fluctuation amplitudes:
\begin{eqnarray}
 \label{eq:Dx_1}
 \tilde{\Delta}_{\Psi^\pm} x_1(t) &=& \sqrt{1 + \frac{1}{\eta} \pm\frac{1}{\sqrt{\eta}}\cos\left(\omega_{\Psi^\pm} t \right)} \\
 \label{eq:Dx_2}
 \tilde{\Delta}_{\Psi^\pm} x_2(t) &=& \sqrt{1 + \frac{1}{\eta} \mp\frac{1}{\sqrt{\eta}}\cos\left(\omega_{\Psi^\pm} t \right)} \\
 \label{eq:Dp_1}
 \tilde{\Delta}_{\Psi^\pm} p_1(t) &=& \sqrt{1 + \eta \pm\sqrt{\eta}\cos\left(\omega_{\Psi^\pm} t \right)} \\
 \label{eq:Dp_2}
 \tilde{\Delta}_{\Psi^\pm} p_2(t) &=& \sqrt{1 + \eta \mp\sqrt{\eta}\cos\left(\omega_{\Psi^\pm} t \right)}
\end{eqnarray}
where the angular frequency
\begin{equation}
 \omega_{\Psi^\pm}=(1-\eta)\omega
\end{equation}
has an absolute value smaller than $\omega$ when the coupling strength $\Omega/\omega$ is in the range $0\leq\Omega/\omega<1$, as shown in figure \ref{fig:frequency_shift}.

\begin{figure}
 \includegraphics[width=\columnwidth,bb = 134 414 478 657,clip]{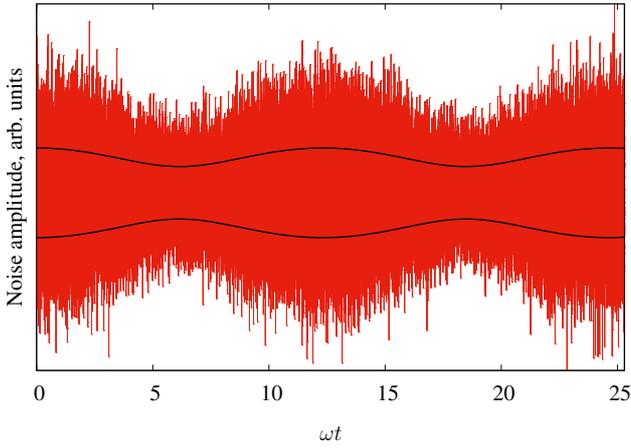}
 \caption{A realization of the $x_1(t)$ fluctuations in the $|\Psi^+ \rangle$ state as a function of $\omega t$. Solid black lines indicate the standard deviation of the fluctuations.}
\label{fig:noise}
\end{figure}

In equations \ref{eq:Dx_1}-\ref{eq:Dp_2},
\begin{eqnarray}
 \tilde{\Delta}_{\Psi^\pm} x_i(t) &=& \frac{\Delta_{\Psi^\pm} x_i(t)}{\sqrt{\hbar/2\omega}} \\
 \tilde{\Delta}_{\Psi^\pm} p_i(t) &=& \frac{\Delta_{\Psi^\pm} p_i(t)}{\sqrt{\hbar\omega/2}}
\end{eqnarray}
are the normalized fluctuation amplitudes. The normalization factors are simply the ground state fluctuation amplitudes $\Delta_0 x$ and $\Delta_0 p$ for a single oscillator. It is worth noticing that the amplitudes of the fluctuations, as measured by the standard deviation of the operators, are time-dependent, which means that the semiaxes of the phase space ellipsoids associated with the entangled states considered in this paper expand and contract periodically. Without coupling ($\Omega/\omega=0$), $\omega_{\Psi^\pm}^\mathrm{nc}=0$ and, therefore, the amplitude of the fluctuations for the $|\Psi^\pm\rangle$ states does not varies with time. With coupling, however, the amplitudes of the fluctuations increase and decrease periodically. In this way, we may think of coupling as causing a kind of dynamical squeezing, which is clearly visible in the realization of the $x_1(t)$ fluctuations in the $|\Psi^+ \rangle$ state displayed in Figure \ref{fig:noise}.

In order to appreciate the effect of coupling on the fluctuations, we indicate that, in the non-coupled case, the amplitude of the fluctuations of the entangled states are
\begin{eqnarray}
\label{eq:nc-boundaries-x_1_p1}
 \frac{\Delta_{\Psi^\pm} x_1^\mathrm{nc}(t)}{\sqrt{\hbar/2\omega}} &=& \sqrt{2\pm 1} = \frac{\Delta_{\Psi^\pm} p_1^\mathrm{nc}(t)}{\sqrt{\hbar\omega/2}} \\
\label{eq:nc-boundaries-x_2_p2}
 \frac{\Delta_{\Psi^\pm} x_2^\mathrm{nc}(t)}{\sqrt{\hbar/2\omega}} &=& \sqrt{2\mp 1} = \frac{\Delta_{\Psi^\pm} p_2^\mathrm{nc}(t)}{\sqrt{\hbar\omega/2}}
\end{eqnarray}
for coordinates and momenta.

\begin{figure}
 \includegraphics[width=\columnwidth,bb = 134 414 478 657,clip]{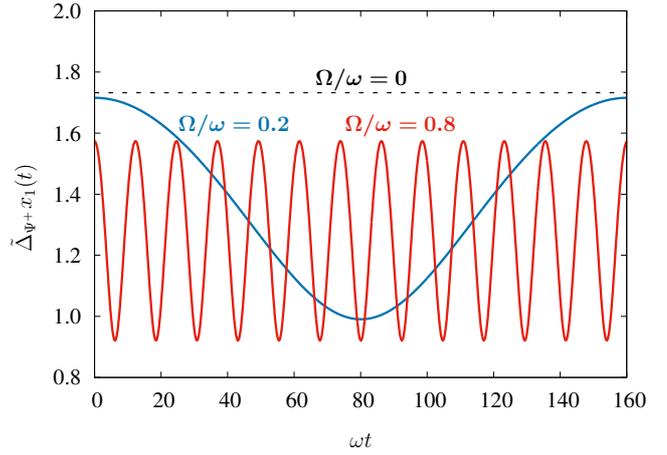}
 \caption{Normalized fluctuation amplitude $\tilde{\Delta}_{\Psi^+} x_1(t)$ as a function of time.}
\label{fig:Delta_x_1_Psi_+}
\end{figure}

Figure \ref{fig:Delta_x_1_Psi_+} shows the normalized fluctuation amplitude $\tilde{\Delta}_{\Psi^+} x_1(t)$ as a function of time for a strong coupling case ($\Omega/\omega=0.2$) and for a very strong coupling case ($\Omega/\omega=0.8$). As the coupling gets stronger, both the maximum and the minimum coordinate fluctuation levels get smaller, that is, coupling leads to a less noisy coordinate.

\begin{figure}
 \includegraphics[width=\columnwidth,bb = 134 414 478 657,clip]{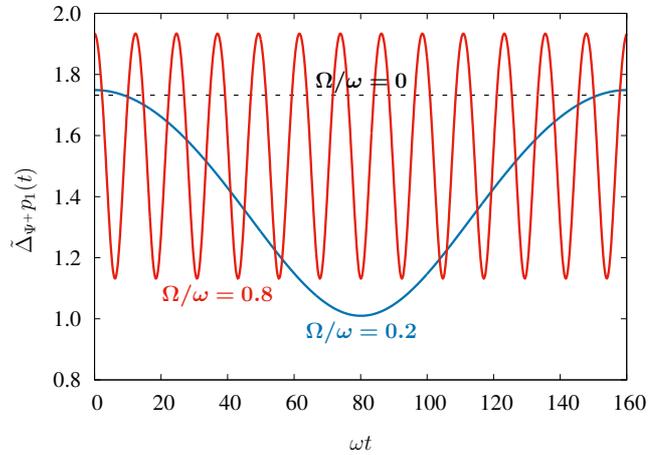}
 \caption{Normalized fluctuation amplitude $\tilde{\Delta}_{\Psi^+} p_1(t)$ as a function of time.}
\label{fig:Delta_p_1_Psi_+}
\end{figure}

The picture is less clear in the case of the momentum fluctuations shown in figure \ref{fig:Delta_p_1_Psi_+}. Although the maximum fluctuation levels are above the uncoupled case, most of the time the fluctuations are below the level of the uncoupled case.

In fact, the normalized coordinate-momentum uncertainty products are summarized by
\begin{eqnarray}
 \label{eq:uncertainty_products}\nonumber
 && \tilde{\Delta}_{\Psi^\pm} x_i(t)\tilde{\Delta}_{\Psi^\pm} p_i(t) = \Bigg(2 + \eta + \frac{1}{\eta} + \cos^2\left(\omega_{\Psi^\pm} t \right) \\ && \mp (-1)^i\times 2\left(\sqrt{\eta} + \frac{1}{\sqrt{\eta}}\right) \cos\left(\omega_{\Psi^\pm} t \right)\Bigg)^{1/2}.
\end{eqnarray}
This should be compared with the non-coupled case, in which the normalized coordinate-momentum uncertainty products are
\begin{eqnarray}
 \label{eq:nc_boundaries-xp_psi_min}
 \Big(\tilde{\Delta}_{\Psi^\pm} x_1(t)\tilde{\Delta}_{\Psi^\pm} p_1(t)\Big)_\mathrm{nc} &=& \sqrt{4 \pm 4 + 1} \\
 \label{eq:nc_boundaries-xp_psi_max}
 \Big(\tilde{\Delta}_{\Psi^\pm} x_2(t)\tilde{\Delta}_{\Psi^\pm} p_2(t)\Big)_\mathrm{nc} &=& \sqrt{4 \mp 4 +1}
\end{eqnarray}
for the $| \Psi^\pm \rangle$ states.

\begin{figure}
 \includegraphics[width=\columnwidth,bb = 134 414 478 657,clip]{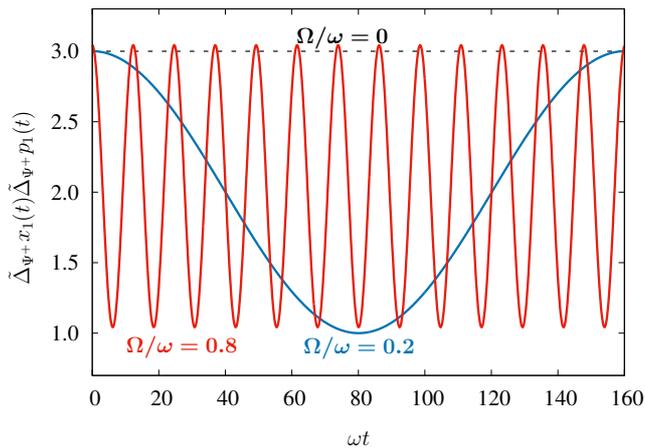}
 \caption{Normalized coordinate-momentum uncertainty product $\tilde{\Delta}_{\Psi^+} x_1(t)\tilde{\Delta}_{\Psi^+} p_1(t)$ as a function of time.}
\label{fig:UP_x_1_p_1_Psi_+}
\end{figure}

\begin{figure}
 \includegraphics[width=\columnwidth,bb = 134 414 478 657,clip]{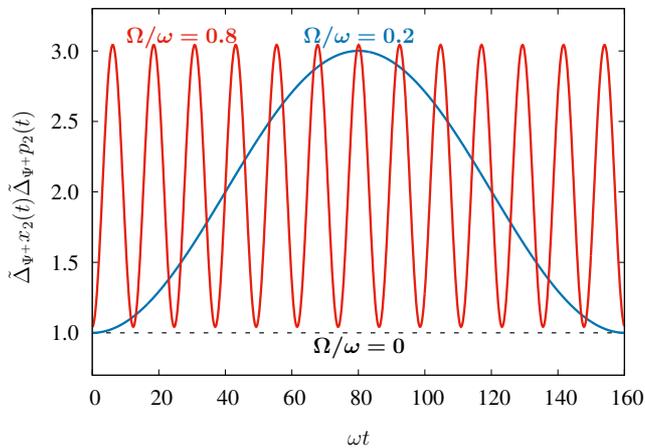}
 \caption{Normalized coordinate-momentum uncertainty product $\tilde{\Delta}_{\Psi^+} x_2(t)\tilde{\Delta}_{\Psi^+} p_2(t)$ as a function of time.}
\label{fig:UP_x_2_p_2_Psi_+}
\end{figure}

Figure \ref{fig:UP_x_1_p_1_Psi_+} illustrates the effect of coupling on the uncertainty product for the entangled state $| \Psi^+ \rangle$. With coupling, the maximum value of the uncertainty product for the $x_1$-$p_1$ pair is slightly above its non-coupled level. However, most of the time, the uncertainty product is well below its non-coupled level, that is, on average, position-position coupling lowered simultaneously both position and momentum fluctuations. On the other hand, figure \ref{fig:UP_x_2_p_2_Psi_+} shows an enhancement of the uncertainty product for the $x_2$-$p_2$ pair of coordinate and momentum owing to coupling. It seems that, for these entangled states, coupling transfers noise from one coordinate-momentum pair to another.

A similar pattern emerges when the state $|\Psi^-\rangle$ is analyzed. This time, however, the uncertainty product for the $x_1$-$p_1$ pair is enhanced above its non-coupled level while the uncertainty product for the $x_2$-$p_2$ pair is, most of the time, well below its non-coupled level. Once again, coupling transferred noise from one coordinate-momentum pair to another.

\section{Conclusion}
\label{sec: conclusion}

We have shown that position-position coupling between quantum oscillators affects the fluctuations of the Bell-like entangled states $|\Psi^\pm\rangle$ that span the single-excitation sector of the hamiltonian. In the uncoupled case, in which $\Omega/\omega=0$, the amplitudes of coordinates and momenta fluctuations are constant in these entangled states. As soon as coupling is turned on, these fluctuation amplitudes become time-varying, with angular frequency $\omega_{\Psi^\pm} = (1-\sqrt{1+2\Omega^2/\omega^2})\omega$. A wide relative variation may be observed in these frequencies since they are almost null at very weak coupling, and they are a sizable fraction of $\omega$ for very strong coupling.

In some cases ($x_1$ and $x_2$ for $|\Psi^+\rangle$ and $|\Psi^-\rangle$, respectively), the fluctuations were reduced below their non-coupled level. In other cases ($p_1$ and $p_2$ for $|\Psi^-\rangle$ and $|\Psi^+\rangle$, respectively), the fluctuations were increased above their non-coupled level.

In the remaining four cases, the amplitudes of the fluctuations cross their non-coupled levels. However, most of the time, the amplitudes are below their non-coupled level for $p_1$ and $p_2$ in states $|\Psi^+\rangle$ and $|\Psi^-\rangle$, respectively, and they are above their non-coupled level for $x_1$ and $x_2$ in states $|\Psi^-\rangle$ and $|\Psi^+\rangle$, respectively. It is found that position-position coupling lowers, on average, the fluctuations of both $x_1$ and $p_1$ ($x_2$ and $p_2$) in the entangled state $|\Psi^+\rangle$ ($|\Psi^-\rangle$) below their non-coupled levels.

Regarding the uncertainty products in the entangled states $|\Psi^\pm\rangle$, coupling acts as a mechanism for noise transfer between one coordinate-momentum pair to another. For instance, in the state $|\Psi^+\rangle$, the $x_1$-$p_1$ uncertainty product is, on average, lowered relatively to its non-coupled level while the $x_2$-$p_2$ uncertainty product is enhanced. Noise transfer mechanisms through coupling may be used to achieve increased precision in entanglement-assisted sensing and metrology.



\begin{thebibliography}{}


\bibitem{Zeilinger:Phys_Scr_1998} A. Zeilinger, Phys. Scr. \textbf{T76}, 203 (1998). 

\bibitem{Horodecki:RMP_2009} R. Horodecki, P. Horodecki, M. Horodecki, K. Horodecki, Rev. Mod. Phys. \textbf{81}, 865 (2009). 

\bibitem{Peschel:BJP_2012} I. Peschel, Braz. J. Phys. \textbf{42}, 267 (2012). 

\bibitem{Aolita+de_Melo+Davidovich:Rep_Prog_Phys_2015} L. Aolita, F. de Melo, L. Davidovich, Rep. Prog. Phys. \textbf{78}, 042001 (2015). 


\bibitem{Dowling+Milburn:PTRSLA_2003} J. P. Dowling, G. J. Milburn, Phil. Trans. R. Soc. London A \textbf{361}, 1655 (2003). 


\bibitem{Jozsa:PRSLA_1998} R. Jozsa, Proc. R. Soc. London A \textbf{454}, 323 (1998). 

\bibitem{Ekert+Jozsa:PTRSLA_1998} A. Ekert, R. Jozsa, Phil. Trans. R. Soc. London A \textbf{356}, 1769 (1998). 

\bibitem{Steane:Rep_Prog_Phys_1998} A. Steane, Rep. Prog. Phys. \textbf{61}, 117 (1998). 

\bibitem{Bennett+DiVincenzo:Nature_2000} C. Bennett, D. DiVincenzo, Nature \textbf{404}, 247 (2000). 

\bibitem{Bouwmeester+Ekert+Zeilinger:PQI_2000} D. Bouwmeester, A. Ekert, A. Zeilinger, The Physics of Quantum Information. Springer, Berlin (2000).

\bibitem{Nielsen+Chuang:QCQI_2000} M. Nielsen, I. Chuang, Quantum Computation and Quantum Information. Cambridge University Press, Cambridge (2000).

\bibitem{Jozsa+Linden:PRSLA_2003} R. Jozsa, N. Linden, Proc. R. Soc. London A \textbf{459}, 2011 (2003). 

\bibitem{Mermin:QCS_2007} N. D. Mermin, Quantum Computer Science. Cambridge University Press, Cambridge (2007).

\bibitem{Preskill:Quantum_2018} J. Preskill, Quantum \textbf{2}, 79 (2018). 


\bibitem{Ekert:PRL_1991} A. Ekert, Phys. Rev. Lett. \textbf{67}, 661 (1991). 

\bibitem{Bennett+Wiesner:PRL_1992} C. H. Bennett, S. J. Wiesner, Phys. Rev. Lett. \textbf{69}, 2881 (1992). 

\bibitem{Bennett+etal:PRL_1993} C. H. Bennett, G. Brassard, C. Crepeau, R. Jozsa, A. Peres, W. K. Wootters, Phys. Rev. Lett. \textbf{70}, 1895 (1993). 

\bibitem{Zukowski+etal:PRL_1993} M. Żukowski, A. Zeilinger, M. A. Horne, A. K. Ekert, Phys. Rev. Lett. \textbf{71}, 4287 (1993). 

\bibitem{Gisin+etal:RMP_2002} N. Gisin, G. Ribordy, W. Tittel, H. Zbinden, Rev. Mod. Phys. \textbf{74}, 145 (2002). 

\bibitem{Gisin+Thew:Nature_Photonics_2007} N. Gisin, R. Thew,  Nature Photonics \textbf{1}, 165 (2007). 

\bibitem{Broadbent+Schaffner:Designs_Codes_Crypto_2016} A. Broadbent, C. Schaffner, Designs, Codes and Cryptography \textbf{78}, 351 (2016). 

\bibitem{Diamanti+etal:npjQI_2016} E. Diamanti, H.-K. Lo, B. Qi, Z. Yuan, npj Quantum Information \textbf{2}, 16025 (2016). 

\bibitem{Yin+etal:Nature_2020} J. Yin, Y.-H Li, S.-K. Liao, M. Yang, Y. Cao, L. Zhang, J.-G. Ren, W.-Q. Cai, W.-Y. Liu, S.-L. Li, R. Shu, Y.-M. Huang, L. Deng, L. Li, Q. Zhang, N.-L. Liu, Y.-A. Chen, C.-Y. Lu, X.-B Wang, F. Xu, J.-Y. Wang, C.-Z. Peng, A. K. Ekert, J.-W. Pan, Nature \textbf{582}, 501 (2020). 

\bibitem{Pirandola+etal:Adv_Opt_Photon_2020} S. Pirandola, U. L. Andersen, L. Banchi, M. Berta, D. Bunandar, R. Colbeck, D. Englund, T. Gehring, C. Lupo, C. Ottaviani, J. Pereira, M. Razavi, J. S. Shaari, M. Tomamichel, V. C. Usenko, G. Vallone, P. Villoresi, P. Wallden, Adv. Opt. Photon. \textbf{12}, 1012 (2020). 


\bibitem{Bollinger+etal:PRA_1996} J. J. Bollinger, W. M. Itano, D. J. Wineland, D. J. Heinzen, Phys. Rev. A \textbf{54}, R4649 (1996). 

\bibitem{Giovannetti+Lloyd+Maccone:Science_2004} V. Giovannetti, S. Lloyd, L. Maccone, Science \textbf{306}, 1330 (2004). 

\bibitem{Giovannetti+Lloyd+Maccone:PRL_2006} V. Giovannetti, S. Lloyd, L. Maccone, Phys. Rev. Lett. \textbf{96}, 010401 (2006). Quantum metrology

\bibitem{Fernholz+etal:PRL_2008} T. Fernholz, H. Krauter, K. Jensen, J. F. Sherson, A. S. S{\o}rensen, E. S. Polzik, Phys. Rev. Lett. \textbf{101}, 073601 (2008). 

\bibitem{Wasilewski+etal:PRL_2010} W. Wasilewski, K. Jensen, H. Krauter, J. J. Renema, M. V. Balabas, E. S. Polzik, Phys. Rev. Lett. \textbf{104}, 133601 (2010). 

\bibitem{Giovanetti+Lloyd+Maccone:Nature_Photonics_2011} V. Giovannetti, S. Lloyd, L. Maccone, Nature Photonics \textbf{5}, 222 (2011). 

\bibitem{Degen+Reinhard+Cappellaro:RMP_2017} C. L. Degen, F. Reinhard, P. Cappellaro, Rev. Mod. Phys. \textbf{89}, 035002 (2017). 


\bibitem{Rempe+Walther+Klein:PRL_1987} G. Rempe, H. Walther, N. Klein, Phys. Rev. Lett. \textbf{58}, 353 (1987). 

\bibitem{Thompson+Rempe+Kimble:PRL_1992} R. J. Thompson, G. Rempe, H. J. Kimble, Phys. Rev. Lett. \textbf{68}, 1132 (1992). 

\bibitem{Brune+etal:PRL_1996a} M. Brune, F. Schmidt-Kaler, A. Maali, J. Dreyer, E. Hagley, J. M. Raimond, S. Haroche, Phys. Rev. Lett. \textbf{76}, 1800 (1996). 

\bibitem{Kimble:Phys_Scr_1998} H. J. Kimble, Phys. Scr. \textbf{T76}, 127 (1998). 

\bibitem{Raimond+Brune+Haroche:RMP_2001} J. M. Raimond, M. Brune, S. Haroche, Rev. Mod. Phys. \textbf{73}, 565 (2001). 

\bibitem{Walther+etal:Rep_Prog_Phys_2006} H. Walther, B. T. H. Varcoe, B.-G. Englert, T. Becker, Rep. Prog. Phys. \textbf{69}, 1325 (2006). 


\bibitem{Bouchiat+etal:Phys_Scr_1998} V. Bouchiat, D. Vion, P. Joyez, D. Esteve, M. H. Devoret, Phys. Scr \textbf{T76}, 165 (1998). 

\bibitem{Nakamura+Pashkin+Tsai:Nature_1999} Y. Nakamura, Yu. A. Pashkin, J. S. Tsai, Nature \textbf{398}, 786 (1999). 

\bibitem{Leibfried+etal:RMP_2003} D. Leibfried, R. Blatt, C. Monroe, D. Wineland, Rev. Mod. Phys. \textbf{75}, 281 (2003). 

\bibitem{Chiorescu+etal:Nature_2004} I. Chiorescu, P. Bertet, K. Semba, Y. Nakamura, C. J. P. M. Harmans, J. E. Mooij, Nature \textbf{431}, 159 (2004). 

\bibitem{Wallraff+etal:Nature_2004} A. Wallraff, D. I. Schuster, A. Blais, L. Frunzio, R.-S. Huang, J. Majer, S. Kumar, S. M. Girvin, R. J. Schoelkopf, Nature \textbf{431}, 162 (2004). 

\bibitem{Blais+etal:PRA_2004} A. Blais, R.-S. Huang, A. Wallraff, S. M. Girvin, R. J. Schoelkopf, Phys. Rev. A \textbf{69}, 062320 (2004). 

\bibitem{Reithmaier+etal:Nature_2004} J. P. Reithmaier, G. Sek, A. L\"{o}ffler, C. Hofmann, S. Kuhn, S. Reitzenstein, L. V. Keldysh, V. D. Kulakovskii, T. L. Reinecke, A. Forchel, Nature \textbf{432}, 197 (2004). 

\bibitem{Clarke+Wilhelm:Nature_2008} J. Clarke, F. K. Wilhelm, Nature \textbf{453}, 1031 (2008). 

\bibitem{Groblacher+etal:Nature_2009} S. Gr\"{o}blacher, K. Hammerer, M. R. Vanner, M. Aspelmeyer, Nature \textbf{460}, 724 (2009). 

\bibitem{Niemczyk+etal:Nature_Physics_2010} T. Niemczyk, F. Deppe, H. Huebl, E. P. Menzel, F. Hocke, M. J. Schwarz, J. J. Garcia-Ripoll, D. Zueco, T. H\"{u}mmer, E. Solano, A. Marx, R. Gross, Nature Physics \textbf{6}, 772 (2010). 

\bibitem{Teufel+etal:Nature_2011} J. D. Teufel, D. Li, M. S. Allman, K. Cicak, A. J. Sirois, J. D. Whittaker, R. W. Simmonds, Nature \textbf{471}, 204 (2011) 


\bibitem{Zeh:FP_1970} H.D. Zeh, Found. Phys. \textbf{1}, 69 (1970). 

\bibitem{Zurek:PRD_1981} W. H. Zurek, Phys. Rev. D \textbf{24}, 1516 (1981). 

\bibitem{Zurek:PRD_1982} W. H. Zurek, Phys. Rev. D \textbf{26}, 1862 (1982). 

\bibitem{Caldeira+Leggett:Physica_1983} A. O. Caldeira, A. J. Leggett, Physica A \textbf{121}, 587 (1983). 

\bibitem{Caldeira+Leggett:Ann_Phys_1983} A. O. Caldeira, A. J. Leggett, Ann. Phys. \textbf{149}, 374 (1983). 

\bibitem{Joos+Zeh:ZPB_1985} E. Joos, H. D. Zeh, Z. Phys. B \textbf{59}, 223 (1985). 

\bibitem{Caldeira+Leggett:PRA_1985} A. O. Caldeira, A. J. Leggett, Phys. Rev. A \textbf{31},1059 (1985). 

\bibitem{Unruh+Zurek:PRD_1989} W. G. Unruh, W. H. Zurek, Phys. Rev. D \textbf{40}, 1071 (1989). 

\bibitem{Brune+etal:PRL_1996b} M. Brune, M. E. Hagley, J. Dreyer, X. Ma\^{\i}tre, A. Maali, C. Wunderlich, J. M. Raimond, S. Haroche, Phys. Rev. Lett. \textbf{77}, 4887 (1996). 

\bibitem{Zurek:RMP_2003} W .H. Zurek, Rev. Mod. Phys. \textbf{75}, 715 (2003). 

\bibitem{Joos+etal:DACWQT_2003} E. Joos, H. D. Zeh, C. Kiefer, D. J. W. Giulini, J. Kupsch, I.-O. Stamatescu, Decoherence and the Appearance of a Classical World in Quantum Theory, Springer, Berlin (2003).

\bibitem{Omnes:BJP_2005} R. Omn\`{e}s, Braz. J. Phys. \textbf{35}, 207 (2005). 

\bibitem{Jacquod+Petitjean:Adv_in_Phys_2009} Ph. Jacquod, C. Petitjean, Advances in Physics \textbf{58}, 67 (2009). 


\bibitem{Paz+Roncaglia:PRL_2008} J. P. Paz, A. J. Roncaglia, Phys. Rev. Lett. \textbf{100}, 220401 (2008). 

\bibitem{Kao+Chou:NJP_2016} J.-Y. Kao, C.-H. Chou, New J. Phys. \textbf{18}, 073001 (2016). 

\bibitem{Makarov:PRE_2018} D. Makarov, Phys. Rev. E \textbf{97}, 042203 (2018). 


\bibitem{Ester+Keil+Narducci:PR_1968} L. E. Estes, T. H. Keil, L. M. Narducci, Phys. Rev. \textbf{175}, 286 (1968). 

\bibitem{Han+Kim+Noz:AmJPhys_1999} D. Han, Y. S. Kim, M. E. Noz, . Am. J. Phys. \textbf{67}, 61 (1999). 

\bibitem{Urzua+etal:SciRep_2019} A. R. Urz\'{u}a, I. Ramos-Prieto, F. Soto-Eguibar, V. Arriz\'{o}n, H. M. Moya-Cessa, Sci. Rep. \textbf{9}, 16800 (2019). 

\bibitem{Macedo+Guedes:J_Math_Phys_2012} D. X. Macedo, I. Guedes, J. Math. Phys. \textbf{53}, 052101 (2012). 

\bibitem{Urzua-Pineda+etal:Q_Rep_2019} A. R. Urz\'{u}a-Pineda, I. Ramos-Prieto, M. Fern\'{a}ndez-Guasti, H. M. Moya-Cessa, Quantum Reports \textbf{1}, 82 (2019). 



\bibitem{Caves:PRD_1981} C.M. Caves, Phys. Rev. D \textbf{23}, 1693 (1981). 

\bibitem{Caves+Schumaker:PRA_1985} C. M. Caves,B. L. Schumaker, Phys. Rev. A \textbf{31}, 3068 (1985). 

\bibitem{Schumaker+Caves:PRA_1985} B. L. Schumaker, C. M. Caves, Phys. Rev. A \textbf{31}, 3093 (1985). 


\bibitem{Heisenberg:ZP_1927} W. Heisenberg, Zeit. Phys. \textbf{43}, 172 (1927). 

\bibitem{Kennard:ZP_1927} E. H. Kennard, Zeit. Phys. \textbf{44}, 326 (1927). 

\bibitem{Condon:Science_1929} E. U. Condon, Science \textbf{69}, 573 (1929). 

\bibitem{Robertson:PR_1929} H. P. Robertson, Phys. Rev. \textbf{34}, 163 (1929). 

\bibitem{Heisenberg:PPQT_1930} W. Heisenberg, The Physical Principles of Quantum Theory. University of Chicago Press, Chicago (1930).







\bibitem{Resch+Lundeen+Steinberg:PRL_2002} K. J. Resch, J. S. Lundeen, A. M. Steinberg, Phys. Rev. Lett. \textbf{88},  113601 (2002). 

\bibitem{Banaszek+etal:PRA_2002} K. Banaszek, A. Dragan, K. W\'{o}dkiewicz, C. Radzewicz, Phys. Rev. A \textbf{66}, 043803 (2002). 

\bibitem{Gubarev+etal:PRA_2020} F. V. Gubarev, I. V. Dyakonov, M. Yu. Saygin, G. I. Struchalin, S. S. Straupe, S. P. Kulik, Phys. Rev. A \textbf{102}, 012604 (2020). 

\bibitem{Jost+etal:Nature_2009} J. D. Jost, J. P. Home, J. M. Amini, D. Hanneke, R. Ozeri, C. Langer, J. J. Bollinger, D. Leibfried, D. J. Wineland, Nature \textbf{459}, 683 (2009). 

\bibitem{Palomaki+etal:Science_2013} T. A. Palomaki, J. D. Teufel, R. W. Simmonds, K. W. Lehnert, Science \textbf{342}, 710 (2013). 



\bibitem{Ockeloen-Korppi+etal:Nature_2018} C. F. Ockeloen-Korppi, E. Damsk\"{a}gg, J.-M. Pirkkalainen, M. Asjad, A. A. Clerk, F. Massel, M. J. Woolley, M. A. Sillanp\"{a}\"{a}, Nature \textbf{556}, 478 (2018). 


\end{thebibliography}
\end{document}